\documentclass[12pt,a4paper]{article}
\pdfoutput=1
\usepackage{jheppub}
\usepackage[all]{hypcap}
\usepackage{color}
\usepackage{graphicx}
\usepackage{epsf}
\usepackage{graphicx,epsfig}
\usepackage{amsmath}
\usepackage{amssymb}
\usepackage{epsfig}
\usepackage{color,colordvi}

\usepackage{amssymb,amsmath,bm}
\usepackage{color}
\usepackage{slashed}
\usepackage{comment}
\usepackage{cancel}
\usepackage[latin1]{inputenc}
\usepackage{amsfonts}
\usepackage{lscape}
\usepackage{amsthm}
\usepackage{booktabs}
\usepackage{array}
\usepackage{rotating}
\usepackage{float}
\usepackage{multirow}
\usepackage{graphicx,rotating,amsmath,multirow,xcolor,colortbl,xcolor}
\usepackage{hyperref,pdflscape,slashed}
\usepackage[capitalise]{cleveref}
\crefformat{section}{Sec.~#2#1#3}

\definecolor{rosso}{cmyk}{0,1,1,0.4}
\definecolor{rossos}{cmyk}{0,1,1,0.55}
\definecolor{rossoc}{cmyk}{0,1,1,0.2}
\definecolor{blu}{cmyk}{1,1,0,0.3}
\definecolor{blus}{cmyk}{1,1,0,0.6}
\definecolor{bluc}{cmyk}{1,1,0,0.1}
\definecolor{verde}{cmyk}{0.92,0,0.59,0.25}
\definecolor{verdec}{cmyk}{0.92,0,0.59,0.15}
\definecolor{verdes}{cmyk}{0.92,0,0.59,0.4}
\definecolor{Gray}{gray}{0.95}

\font\tenrsfs=rsfs10 at 12pt
\font\sevenrsfs=rsfs7
\font\fiversfs=rsfs5
\newfam\rsfsfam
\textfont\rsfsfam=\tenrsfs
\scriptfont\rsfsfam=\sevenrsfs
\scriptscriptfont\rsfsfam=\fiversfs
\def\mathscr#1{{\fam\rsfsfam\relax#1}}

\hypersetup{colorlinks,bookmarksopen,bookmarksnumbered,
linkcolor=blus,pdfstartview=FitH,urlcolor=rossos,citecolor=verde}

\newcommand{\be}{\begin{equation}}
\newcommand{\ee}{\end{equation}}
\newcommand{\bea}{\begin{eqnarray}}
\newcommand{\eea}{\end{eqnarray}}
\newcommand{\beq}{\begin{equation}}
\newcommand{\eeq}{\end{equation}}
\newcommand{\beqa}{\begin{eqnarray}}
\newcommand{\eeqa}{\end{eqnarray}}

\def\ord{\mathcal{O}}

\def\eccoZ{Z_2}

\def\Eq#1{Eq.~(\ref{#1})}

\def\SU{\text{SU} }
\def\U{\text{U} }
\def\HH{H_\mathcal{H}}
\def\hH{h_\mathcal{H}}

\definecolor{colorTC}{rgb}{.2,.7,.2}

\begin{document}

\title{The Hyperbolic Higgs}

\affiliation[a]{Institute of Theoretical Science, University of Oregon, Eugene, OR 97403, USA}
\affiliation[b]{Department of Physics, University of California, Santa Barbara, CA 93106, USA}
\affiliation[c]{Theoretical Physics Department, CERN, Geneva, Switzerland}
 
\author[a]{Timothy Cohen,}
\author[b]{Nathaniel Craig,}
\author[c]{Gian F.~Giudice,}
\author[c]{and Matthew McCullough}

%\emailAdd{tcohen@uoregon.edu}
%\emailAdd{ncraig@physics.ucsb.edu}
%\emailAdd{gian.giudice@cern.ch}
%\emailAdd{matthew.mccullough@cern.ch}

\abstract{We introduce the Hyperbolic Higgs, a novel solution to the little hierarchy problem that features Standard Model neutral scalar top partners.  At one-loop order, the protection from ultraviolet sensitivity is due to an accidental non-compact symmetry of the Higgs potential that emerges in the infrared.  Once the general features of the effective description are detailed, a completion that relies on a five dimensional supersymmetric framework is provided.  Novel phenomenology is compared and contrasted with the Twin Higgs scenario.}

\maketitle

\clearpage
\setcounter{page}{2}
\section{The Hyperbolic Mechanism}
\label{sec:Mechanism}
A zoo of novel top partner states~\cite{Chacko:2005pe,Chacko:2005un,Burdman:2006tz,Poland:2008ev,Chacko:2008cb,Cai:2008au,Craig:2014aea,Batell:2015aha,Curtin:2015bka,Cheng:2015buv,Craig:2016kue,Gherghetta:2016bcc, Kats:2017ojr} have arisen through theoretical explorations of neutral naturalness models~\cite{Chacko:2005pe, Barbieri:2005ri, Chacko:2005vw,  Chacko:2005un, Falkowski:2006qq, Chang:2006ra, Burdman:2006tz, Foot:2006ru, Poland:2008ev,Chacko:2008cb,Cai:2008au,Craig:2013xia, Hedri:2013ina, Craig:2013fga, Craig:2014aea, Geller:2014kta, Burdman:2014zta, Craig:2014roa, Craig:2015pha, Barbieri:2015lqa, Low:2015nqa,  Batell:2015aha, Buttazzo:2015bka, Curtin:2015fna, Cohen:2015gaa,  Curtin:2015bka, Beauchesne:2015lva, Cheng:2015buv,  Bai:2015ztj, Chacko:2015fbc, Craig:2016kue, Yu:2016bku, Yu:2016swa, Barbieri:2016zxn, Gherghetta:2016bcc,  Katz:2016wtw, Cheng:2016uqk, Contino:2017moj, Badziak:2017syq, Kats:2017ojr, Thrasher:2017rpa, Badziak:2017kjk, Serra:2017poj, Ahmed:2017psb, Chacko:2017xpd, Badziak:2017wxn}. The Twin Higgs approach demonstrated the possibility of fermionic top partners that are fully neutral under the Standard Model (SM) symmetries.  However, the state-of-the-art for scalar top partners has not changed since the introduction of Folded Supersymmetry (SUSY)~\cite{Burdman:2006tz}, where the folded stops must be charged under the visible $\SU(2)_W\times \U(1)_Y$ in order for them to couple to the Higgs via the fundamental superpotential.  In this paper, we introduce a new paradigm for neutral naturalness -- the {\it Hyperbolic Higgs} -- which realises the so-far elusive goal (see \emph{e.g.}~Table~1 of~\cite{Curtin:2015fna}) of fully SM neutral scalar top partners.  

The Hyperbolic Higgs mechanism relies on a low-energy scalar potential that exhibits an \emph{accidental} flat direction.  This results from a $\U(2,2)$ non-compact global symmetry of the scalar potential, in contrast to the $\U(4)$ symmetry of the Twin Higgs theory.  Note that $\U(2,2)$ is \emph{not} an accidental symmetry of the low-energy theory: although the scalar potential respects it, the kinetic terms do not.

Explicitly, the scalar potential for the Higgs $H$ (charged under the SM gauge group) and the Hyperbolic Higgs $H_\mathcal{H}$ (charged under an identical copy of the SM gauge group) takes the form
\be
\!\!V_\mathcal{H} = m^2 \Big(\big|H\big|^2 - \big|\HH\big|^2\Big) + \frac{\lambda}{2} \Big( \big|H\big|^2 - \big|\HH\big|^2 \Big)^2 .
\label{eq:scal}
\ee
The classical vacuum manifold is described by a hyperbola (with $f^2\equiv m^2/\lambda$),
\be
\big|\HH\big|^2 - \big|H\big|^2 = f^2 \,.
\ee
This explains the moniker ``Hyperbolic Higgs.''

The flat direction $H_{\rm flat}$ is manifest after the field redefinition
\be
H=H_0 \, \sinh \frac{H_{\rm flat}}{f} \, ,~~~~H_{\mathcal{H}}=H_0 \, \cosh \frac{H_{\rm flat}}{f} \, ,
\label{hyphyp}
\ee
as the potential $V_\mathcal{H}$ depends only on $H_0$ but not on $H_{\rm flat}$.
Since \Eq{hyphyp} corresponds to a hyperbolic rotation but not to a unitary transformation, the form of the kinetic terms is not preserved, signaling that $\U(2,2)$ is not a symmetry of the full theory. On the other hand, we can identify a massless neutral scalar state by expanding the Higgs and Hyperbolic Higgs about their vacuum expectation values using $ H  = (0,v + h/\sqrt{2})$  and $ \HH  = (0,v_{\mathcal{H}} + \hH/\sqrt{2})$. We find two propagating scalar radial modes, $h$ and $\hH$, while the remaining degrees of freedom are eaten by the SM gauge fields and by their Hyperbolic $\SU(2)_\mathcal{H}\times\U(1)_\mathcal{H}$ counterparts. The SM-like Higgs scalar $h_{\text{SM}}$ emerges as the massless state
\be \label{eq:mixing}
h_{\text{SM}} = h \cos \theta  + \hH \sin \theta\,, \qquad\qquad  \tan \theta =  \frac{v}{v_{\mathcal{H}}} ~~.
\ee
While the detailed Hyperbolic phenomenology is sensitive to the UV completion, the above mixing leads to two universal features relevant for Higgs phenomenology: $\cos \theta$ measures the size of an overall modification of SM Higgs couplings, while $\sin \theta$ provides a portal into the Hyperbolic sector contributing to invisible Higgs decays. These features will be described in more detail in \cref{sec:pheno}.

To illustrate the comparison between the Twin and Hyperbolic Higgs models, it is useful to abuse the physics jargon and think of the Higgs as a pseudo-Goldstone of a spontaneously broken non-compact $\U(2,2)$ symmetry (as emphasized above, the analogy holds for the scalar potential, but not for the kinetic terms, implying that some of the usual relations are violated at loop level). 

\begin{figure}[t!]
\centering
\includegraphics[width=0.9\textwidth]{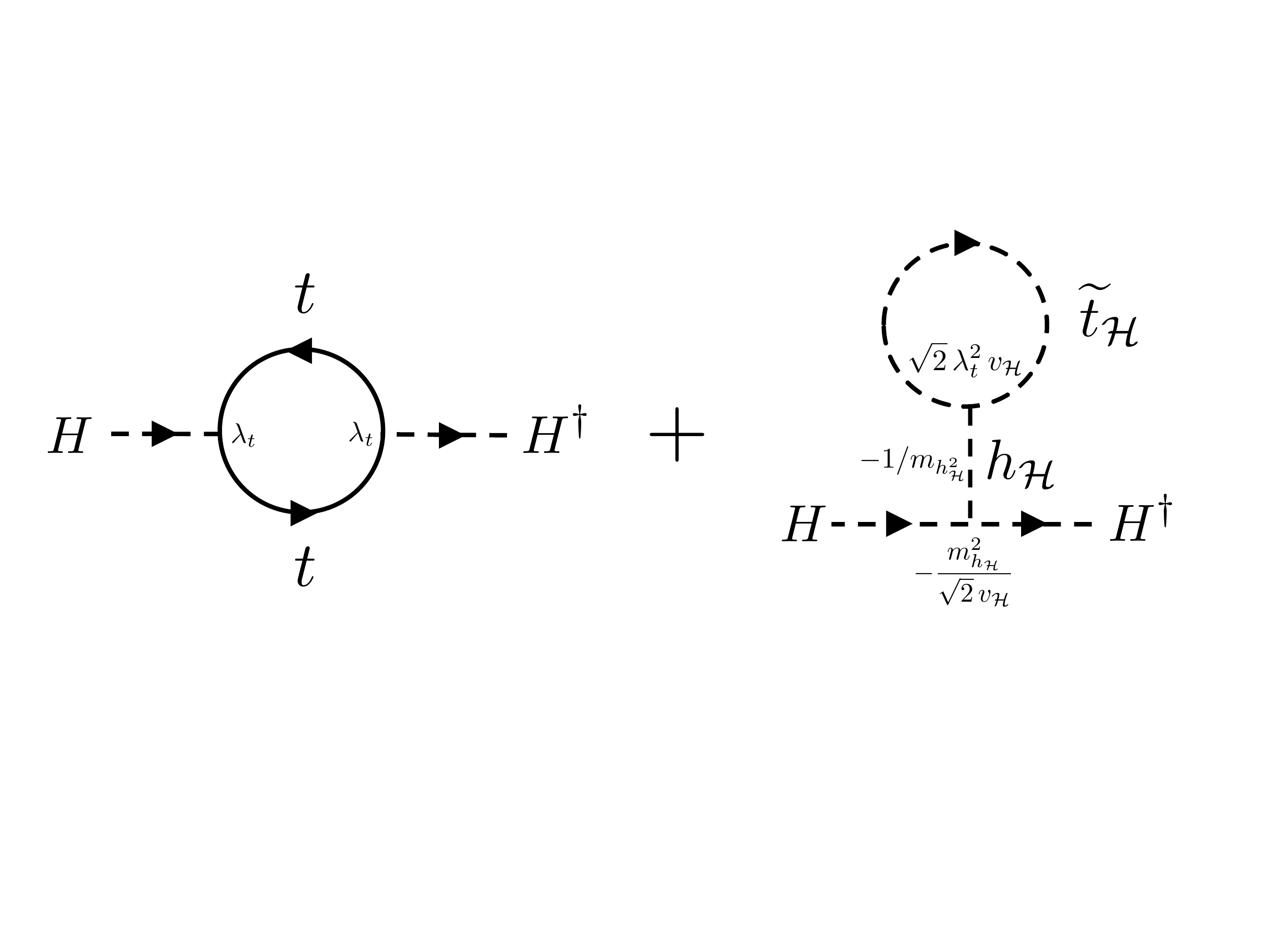}
\caption{The one-loop fixed order corrections to the SM-like Higgs boson mass squared parameter, where $t$ is a SM top quark, and $\widetilde{t}_\mathcal{H}$ is a Hyperbolic stop.  Vertex factors and propagator follow from \cref{eq:scal} and \cref{eq:lag}.}
\label{fig:diag}
\end{figure}

If we are going to solve the little hierarchy problem, we must couple the Higgs sector to matter.  The SM sector has Yukawa couplings involving \emph{fermions} while the Hyperbolic sector contains quartic interactions with \emph{scalars}  
\be
\mathcal{L}  = \left( \lambda_t \,H\, \psi_Q \,\psi_{U^c} + \text{h.c.}\right) 
 + \lambda_t^2  \left( \big|\HH \cdot \widetilde{Q}_\mathcal{H}\big|^2 + \big|\HH\big|^2 \big|\widetilde{U}_\mathcal{H}^c\big|^2 \right) \,,
\label{eq:lag}
\ee
where $\lambda_t$ is the top quark Yukawa coupling, $\psi_Q$ and $\psi_{U^c}$ comprise the SM top quark, and $\widetilde{Q}_\mathcal{H}$ and $\widetilde{U}^c_\mathcal{H}$ are the scalar top partners, in a notation reminiscent of SUSY. An exchange symmetry (properly defined in the following) guarantees the equality of $\lambda_t$ in the two sectors.  Note that in the low-energy effective theory there are no SM charged top partner states.  The field $H$ couples as a SM Higgs, while $\HH$ participates in an interaction analogous to the SUSY Higgs-squark couplings.  Since the UV regulator is equal for both fields due to the exchange symmetry, the one-loop quadratic corrections from these interactions also respect the accidental global $\U(2,2)$ symmetry, $\delta V \propto (\lambda_t/4\pi)^2\, \Lambda^2 \big( \big|\HH\big|^2 - \big|H\big|^2 \big)$, where $\Lambda$ is the UV cutoff.  The opposite sign in the loop corrections for Higgs and Hyperbolic Higgs comes from the replacement of virtual fermions with scalars. This is the central observation of this paper, as this fact guarantees that the one-loop top contributions to the SM Higgs mass squared parameter are insensitive to the UV: the contributions quadratic in the cutoff that result from top quark loops are cancelled by those involving gauge neutral top partner fields.

This UV insensitivity may equivalently be seen by integrating out the heavy radial mode $\hH$ to yield the low energy effective theory
\be
\mathcal{L} = \left( \lambda_t\, H \,\psi_Q\, \psi_{U^c} + \text{h.c.} \right)
 +\, \lambda_t^2\, |H|^2  \Big(\big|\, \widetilde{t}^{\,L}_\mathcal{H}\,\big|^2 +  \big|\,\widetilde{t}^{\,R}_\mathcal{H}\,\big|^2 \Big)\,.
\ee
The procedure is illustrated by the diagrams in \cref{fig:diag} with the $\hH$ line integrated out.  The cancellation of divergences is now manifest, as it works in the same way as in ordinary low-energy SUSY.

The key to the Hyperbolic Higgs framework is that, by moving from a scalar potential with an approximate compact global symmetry $\U(4)$ to an approximate non-compact one $\U(2,2)$, the gauge singlet \emph{fermionic} top partners of the Twin Higgs theory may be replaced by gauge singlet \emph{scalar} top partners.  In other words, the extra minus sign due to the loop of fermions becoming a loop of scalars is compensated by the minus sign due to $\U(4) \rightarrow \U(2,2)$.  

The way we formulated the basic ingredients of the Hyperbolic mechanism makes clear that its most natural realization relies on a SUSY setup. Indeed, in \cref{sec:UV} we introduce a realistic example that utilizes a modification of the mirror structure developed for both SUSY Twin Higgs~\cite{Falkowski:2006qq, Chang:2006ra, Craig:2013fga} and Folded SUSY~\cite{Burdman:2006tz}: the UV theory contains two copies of identical matter content, with a $\eccoZ$ exchange symmetry relating their couplings.  Specifically, the top sector of the model relies on 5D SUSY and a particular choice of Scherk-Schwarz boundary conditions~\cite{Scherk:1979zr,Scherk:1978ta} to lift the unwanted states from the low energy spectrum. In \cref{sec:pheno},  we briefly discuss the novel phenomenology associated with neutral scalar top partners and modified Higgs properties. Our conclusion and a comparison of the Hyperbolic Higgs with other frameworks of neutral naturalness is given in \cref{sec:concl}.  A discussion of how the Hyperbolic Higgs fits within a generic IR framework is given in \cref{app:ir}, and some analytic expressions relevant for computing the symmetry breaking pattern are provided in  \cref{app:santaclausen}.

\section{SUSY 5D Realization of the Hyperbolic Higgs}
\label{sec:UV}
Next we detail a UV model that realises the Hyperbolic Higgs mechanism in the IR.

\subsection{SUSY Breaking Boundary Conditions}
Embedding the SM fields into SUSY multiplets implies that fermions are accompanied by scalars.  Thus to eliminate the SM charged stops from the low energy spectrum, we must appeal to a mechanism that can make the scalars in a supermultiplet heavy without introducing large corrections to the Higgs mass squared parameter.  A similar strategy will be required to lift the fermions in the Hyperbolic sector, leaving only the scalar partners light.  To achieve this, we start with a 5D SUSY model where the extra dimension $y$ is compactified on $S^1/\eccoZ$.  The circumference of this extra dimension is taken to be not too far from the weak scale, $1/(\pi\, R) \sim $ few TeV.  A $\eccoZ$ symmetry is preserved which identifies the points $y \to - y$ and can act non-trivially on the fields that propagate in the bulk of the compactified dimension.

\begin{figure}[t]
\centering
\includegraphics[height=.4\textwidth]{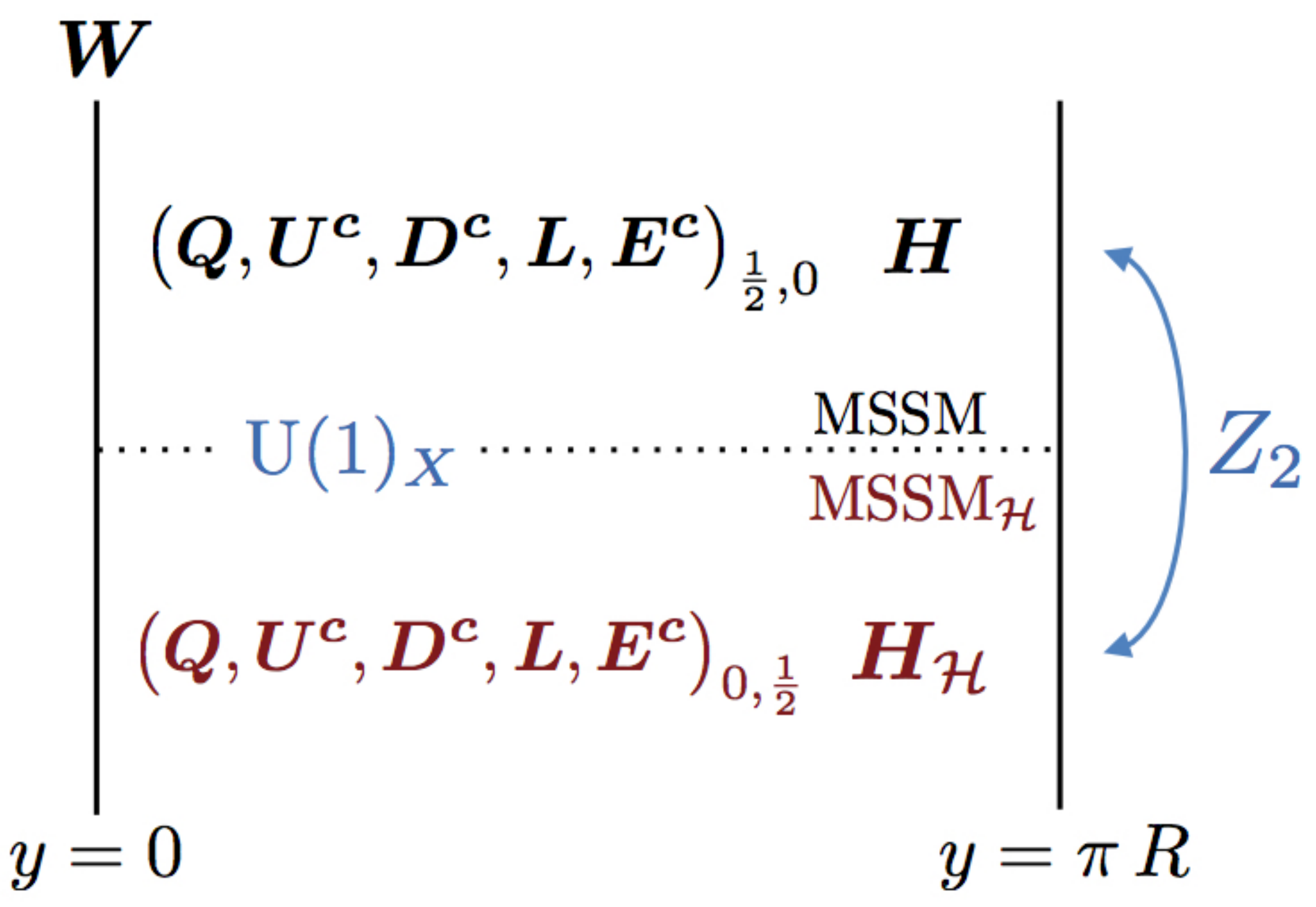}
\caption{A schematic representation of the model.  The bold font signifies full supermultiplets, and the subscript are the Scherk-Schwarz twist parameters $(q_B, q_F)$.  Note that 5D supermultiplets are equivalent to a vector-like pair of 4D supermultiplets.  The Hyperbolic sector is denoted in red.  The new $\U(1)_X$ gauge group gives an interaction between the two sectors and the exchange symmetry $Z_2$ relates coupling constants, as well as interchanging the twists.}
\label{fig:model}
\end{figure}

The Scherk-Schwarz approach is to break SUSY non-locally at the boundaries.  This amounts to a restriction on the extra-dimensional periodicity of the bulk fields, see \emph{e.g.} \cite{Antoniadis:1998sd,Delgado:1998qr,ArkaniHamed:2001mi}.  The 5D field may be decomposed into an infinite tower of 4D mass eigenstates, where the individual modes, whether fermionic or bosonic, have mass $m = (n+q_\phi)/R$, and $q_\phi$ is known as a Scherk-Schwarz twist.

For the theory considered here (shown in \cref{fig:model}), all fields live in the bulk.  The Scherk-Schwarz twists, which determine the zero modes, are denoted with the subscript $(q_B,q_F)$, where $B$ ($F$) is the choice made for the relevant bosons (fermions).  The MSSM matter fields in the visible (Hyperbolic) sector only have fermionic (bosonic) zero modes.  Treating the Scherk-Schwarz parameters as spurions under the exchange MSSM$\,\leftrightarrow\,$MSSM$_{\mathcal{H}}$, the matter field parameters transform as $(q_B,q_F)_{\text{MSSM}} \leftrightarrow (q_F,q_B)_{\text{MSSM}_\mathcal{H}}$, and the non-zero value for these parameters breaks the exchange symmetry.  For concreteness, we take $(q_B, q_F)_{\text{MSSM}} = (1/2,0)$ in the MSSM sector and $(q_B, q_F)_{\text{MSSM}_\mathcal{H}} = (0,1/2)$ in the Hyperbolic sector. Exploring the parameter space as a function of the twist could provide additional model building opportunities, for example as in the Folded SUSY study of~\cite{Cohen:2015gaa}. 

Let us now consider the SM-charged Higgs scalar $H$, interacting with the individual members of the full tower of MSSM top KK modes through a brane-localised superpotential $\bm{W}$ located at $y=0$.  We may calculate the one-loop contribution of the top-stop sector KK tower to the mass of $H$ \`a la Coleman-Weinberg \cite{Coleman:1973jx}, 
\be
\hspace{-5pt}V_\text{CW} (H)  =  \frac{1}{2} \sum_{n} \int \frac{d^4 p}{(2\, \pi)^4}  \left[ \log \frac{p^2+(n+\omega_B^+)^2/R^2}{p^2+(n+\omega_F^+)^2/R^2 } +  \log \frac{p^2+(n+\omega_B^-)^2/R^2}{p^2+(n+\omega_F^-)^2/R^2 } \right]\,,
\label{eq:VCWgeneral}
\ee
where $\omega_{B,F}^\pm = q_{B,F} \pm R\, m_t(H)$.  

Even before performing the integral and summation, Eq.~(\ref{eq:VCWgeneral}) is sufficient to expose the central aspect of this approach to constructing a Hyperbolic Higgs model.  The transformation of the one-loop potential under the exchange $q_B \leftrightarrow q_F$, is $V(H)_{q_B,q_F} = -V(H)_{q_F,q_B}$.  Clearly, the one-loop induced mass terms then flip sign under this exchange.  Ensuring that the one-loop quadratic potential respects the $\U(2,2)$ symmetry is equivalent to the statement that $V(H)$ and $V(\HH)$ are identical up to $H \leftrightarrow \HH$ and the flip in overall sign.   

Regarding the Higgs sector, there are two, equally viable, model-building options: either we keep the full two-Higgs doublet sector, or we adopt a SUSY one Higgs doublet-like structure~\cite{Davies:2011mp, Dimopoulos:2014aua,Dimopoulos:2014psa,Garcia:2015sfa}.  Noting that the essential features of our study are insensitive to this choice, for simplicity we implement the latter.  Our model includes two copies of the MSSM Higgs sector, with boundary conditions such that $N=1$ supermultiplets of Higgs up-type doublets remain in both sectors at low energy.  For the choice of boundary conditions made here, the Higgsinos have a mass $1/(2R)$.  This will lead to a quadratic correction from electroweak loops that we will neglect as it takes the same form and is subdominant to other contributions considered below.  We refer to the zero mode up-type Higgs in the MSSM sector as $H$ and in the Hyperbolic sector as $H_{\mathcal{H}}$.  

\subsection{Matter}
All of the MSSM matter superfields and the complete Hyperbolic copy are included, such that the exchange symmetry MSSM $\leftrightarrow$ MSSM$_\mathcal{H}$ is manifest in the bulk.   Using \cref{eq:VCWgeneral}, we can evaluate the Coleman-Weinberg potential that is generated by the stop-top sector at one loop 
\be
\hspace{-2pt} V_\text{CW}(H) = -\frac{3\, N_c}{32\, \pi^6\, R^4} \Big[ \text{Cl}_5 \big(2\, \pi\, \omega_B^+\big) + \text{Cl}_5 \big(2\, \pi\,  \omega_B^-\big) -\text{Cl}_5 \big(2 \,\pi\, \omega_F^+\big)-\text{Cl}_5 \big(2\, \pi\, \omega_F^-\big)  \Big] \,.
\label{eq:VCWsummed}
\ee
where $N_c =3$ is the number of colors and $\text{Cl}_n(x)$ are Clausen functions which are related to polylogarithms, see \cref{app:santaclausen} for explicit expressions.

The brane-localized Yukawa interaction implies the presence of $\delta(y)$ factors in the equations of motion.  Integrating across the brane to arrive at a self-consistent field profile in the bulk yields an expression for the top mass that depends non-linearly on the Higgs field   
\be
m_t (H) = \frac{1}{\pi\, R} \arctan \Big( \pi\, R \,\lambda_t \,|H|\Big)\, .
\label{eq:mt}
\ee
Thus the top Yukawa coupling $\lambda_t$ is not equal to its SM value $\lambda_t^{\rm SM}\equiv m_t/\langle H \rangle$, as we will discuss in \cref{sec:pheno}.

For illustration, we can expand $V_\text{CW}^\text{MSSM} + V_\text{CW}^{\text{MSSM}_\mathcal{H}}$ to quadratic order in the unbroken phase, to expose that the scalar zero modes do not remain massless at one-loop:
\begin{align}
V_\text{CW} =   - \frac{21 \,\zeta (3)\, \lambda_t^2}{32\, \pi^2\, (\pi\,R)^2}  \bigg\{& N_c  \left( \big|H\big|^2- \,\big|\HH\big|^2 \right)   -   \big|\widetilde{Q}_\mathcal{H}\big|^2 - 2\,  \big|\widetilde{U}_\mathcal{H}^c\big|^2  \bigg\}\, + \dots\,,
\label{eq:soft}
\end{align}
where $\widetilde{Q}_\mathcal{H}$ and $\widetilde{U}_\mathcal{H}^c$ are the Hyperbolic stop fields, and $\zeta(x)$ is the Riemann zeta function.  This manifestly respects the $\U(2,2)$ symmetry, demonstrating the expected result that all quadratic sensitivity to the cutoff obeys this approximate symmetry.  Note, however, that the contributions prefer a vev for $H$ rather than $H_\mathcal{H}$.  Thus, although $\U(2,2)$-symmetric, these terms alone are not sufficient to achieve electroweak symmetry breaking.  This will be remedied in \cref{sec:gauge}.

\subsection{Gauge}
\label{sec:gauge}
For minimality in the $\SU(3)_c\times\SU(2)_W\times\U(1)_Y$ gauge sectors, we choose boundary conditions such that a full $N=1$ vector supermultiplet for the $\SU(2)_W\times\U(1)_Y$ gauge fields survives in the IR, whereas we will keep only zero modes for $\SU(3)_c$ gauge bosons.  This lifts the gluinos such that LHC bounds on their production are not relevant.  In principle, the heavy Hyperbolic gluinos lead to corrections for the Hyperbolic stop mass which ultimately feeds into the Higgs mass squared parameter, but in practice these are subdominant to the corrections already considered here.    

Critical to the structure of this UV-completion is the introduction of a new $\U(1)_X$ gauge symmetry.  By charging the Higgs fields as $Q_X (H,H_\mathcal{H}) = (1,-1)$, the associated $D$-term potential will provide the $\U(2,2)$-invariant quartic interaction.  The gauge symmetry must be spontaneously broken, leading to freedom in choosing charges for the matter fields.  We will assume that the $\U(1)_X$ matter charge assignment is at least consistent with the top Yukawa interaction, while other Yukawa interactions can arise as higher dimension operators involving the $\U(1)_X$-breaking field if necessary.  If any matter fields are brane-localised, we assume that their charge assignments do not generate mixed $\U(1)_X\text{ -- }\SU(3)_c$ anomalies.   If required, mixed electroweak anomalies can be removed with the addition of anomalon fields.

We impose boundary conditions that yield a massless $\U(1)_X$ gauge boson and gaugino, consistent with 4D $N=1$ SUSY.  This symmetry is then spontaneously broken non-supersymmetrically by an additional scalar field, thereby splitting the massive vector supermultiplet.  We also assume that the brane-localised Fayet-Iliopoulos (FI) $D$-term is non-zero (this is captured by $f_X$ in \cref{eq:U1HQuartic} below).  This allows us to flip the sign of the $\U(2,2)$ preserving soft-masses in \cref{eq:scal} such that the vev of $H_{\mathcal{H}}$ is non-zero.  Putting it all together, the $\U(1)_X$-induced quartic interaction from integrating out the auxiliary fields is 
\be
V_{\U(1)_X} = \frac{g_X^2}{2} \,\xi\, \Big( |H_\mathcal{H}|^2- |H|^2 - f_X^2 \Big)^2 \,,
\label{eq:U1HQuartic}
\ee
and the $D$-term non-decoupling factor is~\cite{Batra:2003nj, Maloney:2004rc, Cheung:2012zq} 
\be
\xi = \left(1- \frac{M_X^2}{M_S^2} \right)\,,
\ee
where $M_X$ is the vector boson mass and $M_S$ is the physical mass of the scalar in the massive vector multiplet.  A splitting between $M_X$ and $M_S$ can be achieved with brane-localized SUSY breaking, and this will lead to the mass correction for $H$ and $\HH$ given in \Eq{eq:SUSYbreak} below.

Due to the SM Higgs charge assignment, electroweak and $\U(1)_X$ breaking cause the $Z_X$ boson to mix with the SM $Z$ boson at tree level.  This shifts $M_Z$ with respect to its SM value, changing the value of the $\rho$ parameter as
\be
\Delta \rho =  \frac{4\,g_X^2\, M_W^2}{g^2\, M_X^2} ~~\Rightarrow~~ \frac{M_X}{g_X} >  8.6~{\rm TeV} \,,
\label{eq:DeltaRho}
\ee
where $M_W$ is the SM $W$-boson mass.  This expression has been expanded to leading order in $M_W^2/M_X^2$, and the limit is obtained by taking the $95\%$ CL allowed band for $\Delta \rho = (3.7 \pm 2.3)\times 10^{-4}$~\cite{Patrignani:2016xqp}.

\subsection{Higgs}
Without loss of generality, we only consider the neutral component of the Higgs fields.   The full scalar potential used in our analysis is 
\be
V = V_\text{CW} (H) + V_\text{CW} (H_\mathcal{H}) + V_{\U(1)_X} + V_{\cancel{\U(2,2)}} \,,
\label{fullscalar}
\ee 
where $V_\text{CW} (H)$ and $V_\text{CW} (H_\mathcal{H})$ capture loops from the top-stop sector as given in \cref{eq:VCWsummed}, $V_{\U(1)_X}$ is given in \cref{eq:U1HQuartic}, and the final $\U(2,2)$-violating contribution to the scalar potential included in our analysis is  
\begin{align}
V_{\cancel{\U(2,2)}}  = & \big(\widetilde{m}^2 + \widetilde{m}_X^2\big)  \Big(\big|H\big|^2+ \big|H_{\mathcal{H}}\big|^2 \Big)  + \frac{g_Z^{2}}{2}  \Big( \big|H\big|^4+  \big|H_{\mathcal{H}}\big|^4 \Big)\,.
\label{eq:break}
\end{align}
The quartic interaction in \cref{eq:break} is the usual MSSM $D$-term and its Hyperbolic counterpart, with $g_Z^{2} = (g^2 + g^{\prime\, 2})/4 = M_Z^2/(2\, v^2)$ and $v\simeq 174$ GeV. Moreover, $\widetilde{m}^2$ is a brane-localised soft term, which is the tunable parameter needed to achieve the desired SM-like vacuum,\footnote{Any combination of two soft masses can be written as a symmetric mass plus an asymmetric mass.  The latter can be absorbed into the brane-localised FI-term of \cref{eq:U1HQuartic}, implying that the full set of possible brane-localised soft masses is captured by assuming $\widetilde{m}$ respects the $\eccoZ$ symmetry.} and $\widetilde{m}_X^2$ is the loop-induced $\U(1)_X$ contribution to the soft mass discussed next.

Brane-localised SUSY breaking is required to realise non-decoupling $D$-terms.  Therefore, there is a one-loop $\U(2,2)$-violating soft mass for the Higgs and Hyperbolic Higgs~\cite{Intriligator:2010be} 
\be
\widetilde{m}_X^2  = \frac{g_X^2\,M_X^2}{16\, \pi^2}  \log\frac{(1-\xi/2)^4}{(1-\xi)^3}  ~,
\label{eq:SUSYbreak}
\ee
in the $R$-symmetric limit.  Note that there is a subdominant correction from loops involving Higgsinos that we neglect for simplicity.  Requiring consistency with the $\Delta \rho$ constraint given in \cref{eq:DeltaRho} results in an irreducible correction to the soft masses for both the Higgs and its Hyperbolic partner
\be
\widetilde{m}_X^2 \gtrsim \big(440 \text{ GeV}\big)^2 \left(\frac{g_X}{0.8}\right)^4  \log \frac{(1-\xi/2)^4}{(1-\xi)^3}   \,.
\label{eq:softapprox}
\ee
This contribution highlights a central tension in the model.  Taking $g_X$ or $\xi$ small will reduce the constraint on $\widetilde{m}_X^2$; however, as either parameter goes to zero, the quartic interaction required to achieve the Hyperbolic Higgs vacuum (see \cref{eq:U1HQuartic} above) also vanishes.  Noting that this SUSY-breaking contribution is significantly smaller than the compactification scale, we are justified in omitting it when performing the sum over KK mode masses that yields \cref{eq:VCWsummed}.

To explore spontaneous electroweak and Hyperbolic symmetry breaking, we shift the vacuum to $H_{(\mathcal{H})} = v_{(\mathcal{H})}+h_{(\mathcal{H})}/\sqrt{2}$, where $v_{(\mathcal{H})}$ is the vev and $h_{(\mathcal{H})}$ is the physical excitation of $H_{(\mathcal{H})}$.  The complete expressions relevant for the conditions of electroweak breaking, which are used for the numerical analysis, are given in \cref{app:santaclausen}.

We can gain some intuition for how the relevant terms scale by taking the Hyperbolic limit
\be
g_X^2\, \xi \gg g_Z^2 
\label{eq:hyplim}
\ee
and expanding in the dimensionless parameters $\alpha_t$ and $\alpha_{t_\mathcal{H}}$, defined as  
\be
\alpha_{t_{(\mathcal{H})}} \equiv \pi\, R\, m_t \big(v_{(\mathcal{H})}\big)\,,
\label{eq:alphaDef}
\ee 
where $m_t(v)$ is given in \cref{eq:mt}.  Remarking that this expansion can break down for some values of $v_\mathcal{H}$ (see \cref{app:santaclausen} for a discussion), we find the following relationship between the soft masses and the vev arising from to  explicit $\U(2,2)$-breaking contributions
\be
\widetilde{m}^2 + \widetilde{m}_X^2  \simeq  \frac{v_\mathcal{H}^2}{2}  \left( \frac{N_c \,\lambda_t^4}{24\, \pi^2}\, \Big[ 11 + 21\, \zeta(3) - 6 \log \big( \lambda_t \,v_\mathcal{H}\, \pi\, R \big) \Big] - g_Z^2 \right) \, .
\label{eq:moresoft}
\ee
We conclude that the total soft mass is parametrically suppressed with respect to $v_\mathcal{H}$ by either a loop factor (the one-loop stop-top corrections to the quartic) or a small parameter (the electroweak gauge coupling from the $D$-term quartic).  Numerically, however, the former is only a suppression of $\sim 0.3$.  The second minimization condition determines the FI term to be
\be
f_{X}^2 \simeq \left( 1+ \frac{g_Z^2}{2\,g_X^2\, \xi}\right) v_\mathcal{H}^2 +
\frac{21\, N_c\, \zeta(3)\, \lambda_t^2}{32\, \pi^4\, R^2\, g_X^2\, \xi} \,.
\label{eq:fXMin}
\ee
In order to avoid fine-tuning the $\U(2,2)$-symmetric terms, the second term should be subdominant.

Taking the Hyperbolic limit given in \Eq{eq:hyplim}, the SM-like Higgs mass is
\be
m_h^2  \simeq  (2-\Omega)\, M_Z^2 +\frac{N_c\, \lambda_t^4 }{2\, \pi^2}\, v^2 \log \frac{v_\mathcal{H}}{v}  \, ,
\label{eq:higgsmass}
\ee
where $\Omega$ is defined in \cref{app:santaclausen} and has the useful property that it vanishes in the Hyperbolic limit. Note that, in the Hyperbolic limit, the first term in \cref{eq:higgsmass} exhibits an additional factor $2$ when compared to the MSSM Higgs boson tree-level mass, as has also been noted for the SUSY Twin Higgs~\cite{Craig:2013fga}.  The parameter $\Omega$ essentially measures the deviations from the Hyperbolic limit and \Eq{eq:higgsmass} reveals that, since $\sqrt{2} M_Z = 129$ GeV, some departure from the Hyperbolic limit is necessarily required to realise $m_h = 125$ GeV, and this deviation must grow as $v_\mathcal{H}/v$ grows.   The second term in \cref{eq:higgsmass} is a universal IR effect from running in the SM effective theory from the new-physics scale to the weak scale.  Noting that the ratio of the Hyperbolic stop mass to the top mass is $m_{\tilde{t}_\mathcal{H}}/m_t \simeq v_\mathcal{H}/v$, this term is in complete analogy with the usual MSSM expression, with Hyperbolic stop squarks replacing the role of the usual colored stop squarks.

In conclusion, the three relevant scales of our model are the UV cutoff $\Lambda$ where the 5D SUSY description becomes strongly coupled ($\Lambda \sim 1/R$), the physical mass of the Hyperbolic Higgs boson ($m_{h_\mathcal{H}} \sim g_X v_\mathcal{H}$), and the physical mass of the SM-like Higgs boson ($m_h \sim g_Z v$).  Due to the different parametric dependences the desired separation of scales $\Lambda \gg m_{h_\mathcal{H}} \gg m_h$ is possible without excessive fine-tuning.  To substantiate this claim, we turn to a more detailed analysis of the tuning next.

\subsection{Tuning}
Let us consider the tuning required for the input parameters $\tilde{m}^2$ and $f_X^2$ when they satisfy the vacuum minimization conditions \cref{eq:moresoft} and \cref{eq:fXMin} respectively. We consider the following standard measure of fine-tuning~\cite{Barbieri:1987fn}
\be
\Delta = \sqrt{ \left( \frac{d \log v^2}{d \log \widetilde{m}^2} \right)^2 + \left( \frac{d \log v^2}{d \log f_X^2} \right)^2  } \, ,
\ee
which gives the tuning in input parameters required to realise the observed weak scale. 

Using the derivative chain rule, we can write
\beq
\frac{d \log v^2}{d \log \widetilde{m}^2} =\frac{ 2\, \widetilde{m}^2 \left( 1 + \tan^2 \alpha_t \right)}
{\tan \alpha_t \left(  \frac{\partial \widetilde{m}^2}{\partial \alpha_t}+ \frac{\partial \widetilde{m}^2}{\,\,\partial \alpha_{t_\mathcal{H}}}\left. \frac{\,\,d \alpha_{t_\mathcal{H}}}{d \alpha_t}   \right|_{f_X^2~{\rm fixed}}\right) } \, ,
\eeq
where $\alpha_{t_{(\mathcal{H})}}$ is defined in \cref{eq:alphaDef}.
The derivatives in the denominator can be readily evaluated using the expressions in Eqs.~(\ref{eqmin1})--(\ref{eqmin2})
and the second term in parenthesis is necessary to properly estimate the fine-tuning, since for a fixed value of $f_X^2$ a change in $\alpha_t$ implies a correlated change in $\alpha_{t_\mathcal{H}}$.  The same procedure is used to compute the impact of a variation in $f_X^2$ for fixed $\widetilde{m}^2$.

For large $\widetilde{m}_X$ this expression is approximately given by
\be
\Delta \simeq 4\, \frac{ \widetilde{m}_X^2}{m_h^2} \,.
\ee
Using \Eq{eq:softapprox}, we find a limit on the fine-tuning when $m_h = 125$ GeV of
\be
\Delta \gtrsim 50 \left(\frac{g_X}{0.8}\right)^4  \log\frac{(1-\xi/2)^4}{(1-\xi)^3}   \,.
\ee
This demonstrates to what extent this UV-completion can capture the Hyperbolic limit, since increasing $g_X$ will increase this source of fine-tuning.  This is not surprising, since any quasi-flat direction will be lifted once SUSY is broken, and it is necessary to use SUSY breaking to keep the $\U(1)_X$ $D$-term potential from decoupling while also lifting the extra gauge boson beyond the limit on the $\rho$ parameter.  A natural weak scale can only be one loop factor below the $\U(1)_X$ sector SUSY breaking scale.

\begin{figure}[t!]
\centering
\includegraphics[height=.39\textwidth]{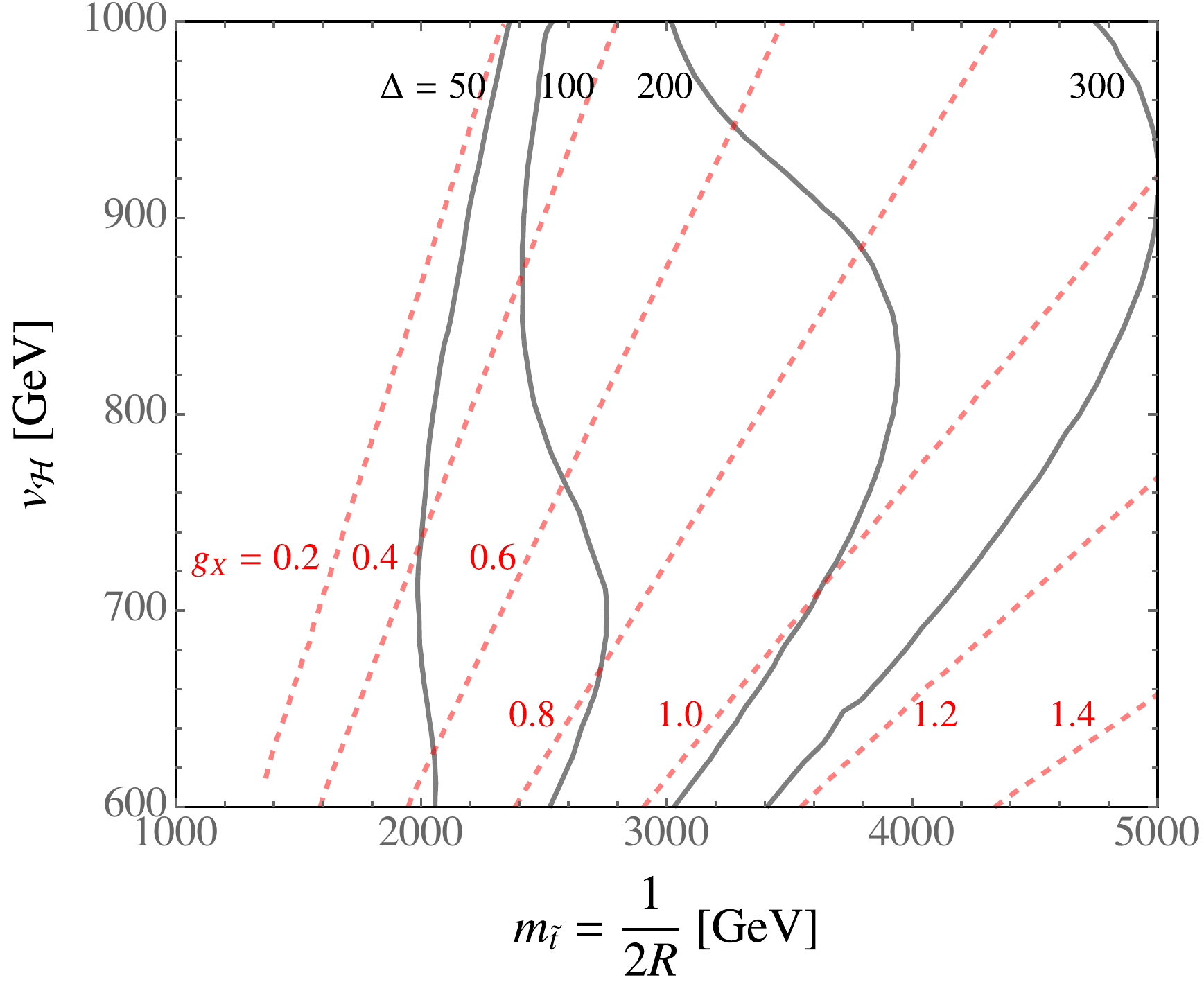} ~ \includegraphics[height=.39\textwidth]{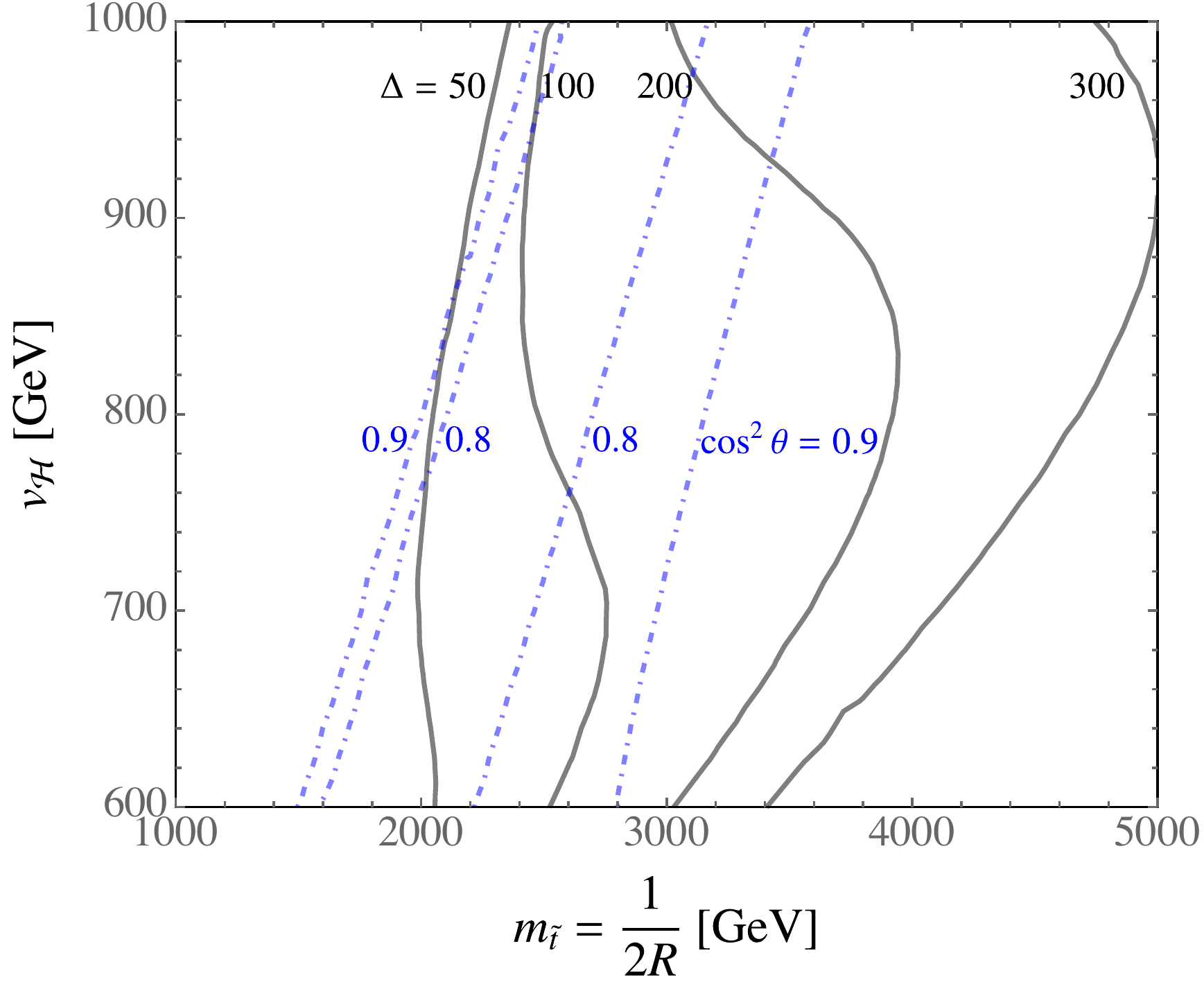}
\caption{Left panel: in the Hyperbolic spontaneous symmetry breaking scale ($v_\mathcal{H}$) vs colored stop mass ($m_{\tilde{t}}$) plane, contours of the fine-tuning parameter $\Delta$ (solid grey), alongside contours of the required value of $g_X$ (dashed red) to realise a leading order prediction of $m_h = 125$ GeV, taking $\xi = 0.6$ and large $\tan\beta$. Right panel: the suppression factor ($\cos^2 \theta$) of SM-like Higgs signal strengths due to mixing (dot-dash blue), under the same assumptions of the previous panel.}
\label{fig:tuning}
\end{figure}

Using the complete expressions given in \cref{app:santaclausen}, in the left panel of \cref{fig:tuning} we show the required value of $g_X$ to realise a physical SM-like Higgs mass of $125 \text{ GeV}$,  alongside the fine-tuning parameter $\Delta$ as a function of the Hyperbolic spontaneous symmetry breaking scale ($v_\mathcal{H}$) and MSSM colored stop mass ($m_{\tilde{t}}$). This figure corresponds to the case of large $\tan \beta$ for a two-Higgs doublet model. Varying $\tan \beta$ gives further flexibility in the allowed choices for $g_X$, $v_\mathcal{H}$ and $m_{\tilde{t}}$. We see that, for large $\tan \beta$, the observed Higgs mass can be realised for colored stops around 2--3 TeV and fine-tuning at the $\sim 1\%$ level.  Increasing the gauge coupling $g_X$ increases the Hyperbolic quartic and thus allows for heavier colored stop squarks.  However, one cannot increase the quartic arbitrarily due to the soft mass contribution of \Eq{eq:SUSYbreak}, which comes to dominate the fine-tuning as $g_X$ becomes large.  This limits the extent to which the Hyperbolic scenario can relieve the fine-tuning.  One is also pushed to regions of larger fine-tuning by requiring that modifications of the Higgs signal strength from Higgs portal mixing remain small.

This discussion also neglects possible tuning associated with aspects of the 5D setup such as radius stabilisation and UV-dependent brane-localized kinetic terms.  If the Scherk-Schwarz setup is strictly necessary to realise Hyperbolic Higgs models in the IR, then any fundamental sources of fine-tuning that are a consequence of working with a 5D SUSY scenario are endemic to the Hyperbolic Higgs.

\newpage
\section{Phenomenology}
\label{sec:pheno}

The phenomenological possibilities for Hyperbolic Higgs models are broad. As with other models of neutral naturalness, the principal signatures involve the Higgs sector.

Much as in Twin Higgs models, the tree-level mixing between the SM and Hyperbolic Higgs doublets leads to a universal reduction in Higgs couplings ($y_{hPP}$) compared to SM predictions ($y_{hPP}^{\rm SM}$)
\beq
\frac{y_{hPP}}{y_{hPP}^{\rm SM}} -1= \cos \theta -1 \simeq -\frac{\omega^2\, v^2}{2\, v_{\mathcal{H}}^2}
\simeq -\omega^2 \left( \frac{\rm TeV}{v_{\mathcal{H}}} \right)^2 \, 1.5\% \, ,
\eeq
where $\omega$ is a model-dependent coefficient, equal to one in the Hyperbolic limit and whose expression in our proposed model in given in \Eq{omeghino}. The numerical value of the suppression with respect to SM rates, $\cos^2 \theta = ({y_{hPP}}/{y_{hPP}^{\rm SM}})^2$, is shown in the right panel of \cref{fig:tuning}.

Such a universal reduction in Higgs couplings drops out of Higgs branching ratios, but can be detected in measurements of the Higgs width and, more importantly, of Higgs production rates. Current Higgs coupling measurements (namely the combined ATLAS + CMS 7+8 TeV $2 \sigma$ bounds on modifications to Higgs production rates \cite{Khachatryan:2016vau}) require $v_\mathcal{H} \gtrsim 500$ GeV. This is ultimately a weaker constraint on the natural parameter space of the Hyperbolic Higgs than that imposed by precision electroweak limits on $\U(1)_X$.  As larger values of $v_\mathcal{H}$ are allowed without significantly increasing fine-tuning, future LHC Higgs coupling measurements are unlikely to place strong limits on the model.

Because of the non-linearities in \Eq{eq:mt}, the modification of the Higgs coupling to top receives additional non-universal corrections. Taking the first derivative of $m_t (H)$ in \Eq{eq:mt} around $H=v +h_{\rm SM} \cos \theta /\sqrt{2}$, we find
\bea
\frac{y_{htt}}{y_{htt}^{\rm SM}} -1&=& \frac{\cos \theta}{1+\tan^2\alpha_t} -1 \simeq -\frac{\omega^2\, v^2}{2\, v_{\mathcal{H}}^2}-\pi^2\, R^2\, \lambda_t^2\, v^2 
\nonumber \\[8pt]
&\simeq& -\omega^2 \left( \frac{\rm TeV}{v_{\mathcal{H}}} \right)^2 \, 1.5\% -
\left( \frac{5~{\rm TeV}}{1/R} \right)^2 \, 1.2\% \, .
\eea

Mixing between the SM and Hyperbolic Higgs doublets also induces corrections to the Higgs self-coupling ($y_{h^3}$) relative to SM predictions ($y_{h^3}^{\rm SM}$)
\bea
\frac{y_{h^3}}{y_{h^3}^{\rm SM}} -1&=& \cos^3 \theta + \frac{v}{v_{\mathcal{H}}} \sin^3 \theta -1 \simeq - \frac{3\, \omega^2\, v^2}{2\, v_{\mathcal{H}}^2} \simeq - \omega^2 \left( \frac{\rm TeV}{v_{\mathcal{H}}} \right)^2 \, 4.5\% \, .
\eea
This analytic expression is exact, once we neglect the tower of higher-dimensional operators in the CW potential. 

Mixing between the SM and Hyperbolic Higgs also imbues the SM-like Higgs with couplings to light states in the Hyperbolic sector.  In contrast to the Twin Higgs, there are no light chiral fermions in the Hyperbolic sector, while the Hyperbolic scalars all acquire soft masses greater than, or comparable to, the electroweak scale. Consequently, the only possible light Hyperbolic degrees of freedom beneath the electroweak scale are the Hyperbolic photon and gluon. The effective coupling to Hyperbolic photons leads to a vanishingly small contribution to the Higgs invisible decay rate, at least an order of magnitude smaller than the SM contribution from $h \rightarrow 4\, \nu$. The coupling to Hyperbolic gluons, on the other hand, is more promising due to the effects of Hyperbolic confinement. Neglecting SUSY breaking corrections to Hyperbolic stop masses and their possible mixing (which is a good approximation as long as $1/R \lesssim 10 \, v_{\mathcal{H}}$), we find
\beq
{\rm BR}(h_{\rm SM} \to g_\mathcal{H}\,g_\mathcal{H}) = \frac{\sin^2\theta}{4} \frac{ \alpha_{s,\mathcal{H}}^2}{\alpha_s^2} \frac{v^2}{v_{\mathcal{H}}^2} \, {\rm BR}(h_{\rm SM} \to g\,g)
\simeq \omega^2 \left( \frac{\rm TeV}{v_{\mathcal{H}}} \right)^4 \, 2\times 10^{-5}\,,
\eeq
where the factor $1/4$ comes from the ratio of the loop functions (fermionic top for the SM and scalar stops for the Hyperbolic sector) and the factor $v^2/v_{\mathcal{H}}^2$ comes from the ratio between the scales of the dimension-5 operators mediating the decay. We have also kept explicitly the factor $\alpha_{s,\mathcal{H}}^2 / \alpha_s^2$ to account for different renormalisation effects in the SM and Hyperbolic sectors.

Assuming that there is no additional spontaneous symmetry breaking in the Hyperbolic sector, the phenomenology of these Hyperbolic gluons is much like those in Folded SUSY~\cite{Curtin:2015fna} or the fraternal Twin Higgs~\cite{Craig:2015pha}. Once produced, the Hyperbolic gluons form Hyperbolic glueballs, the lightest of which have $J^{PC} = 0^{++}$ and decay back to the SM via Higgs mixing, with a proper length on the order of meters to kilometers. In contrast to Folded SUSY and the fraternal Twin Higgs, this is the only process leading to the production of Hyperbolic glueballs from SM states, as there are no open Higgs decay modes or Drell-Yan production channels involving additional states charged under Hyperbolic color. In this respect, the Hyperbolic Higgs is the most predictive model of neutral naturalness in terms of the SM production rate for dark glueball states. As with the fraternal Twin Higgs, however, the radial Hyperbolic Higgs mode provides an additional portal into the Hyperbolic sector. 

The above glueball phenomenology assumes no additional spontaneous symmetry breaking in the Hyperbolic sector. However, the Hyperbolic stops may obtain a vacuum expectation value if the vacuum structure favors it. In this case  Hyperbolic QCD and electromagnetism can be spontaneously broken, removing the light gauge bosons to leave no new light states below the weak scale. Not only does this significantly reduce collider and cosmological signatures, but it also raises the intriguing prospect that $2/3$ of the top partner degrees of freedom would be eaten, hence the longitudinal modes of the Hyperbolic gluons would also be top partners in this picture.  The other $1/3$ of the top partner degrees of freedom would also mix with the Higgs boson, implying that {\it the Higgs is partially its own top partner}~\cite{Grandpaw}.

The phenomenology of the gauge neutral Hyperbolic sfermions may be varied.  They may be stable, due to an unbroken $Z_{2R}$ symmetry.  In this case the top partners may make up some part of the dark matter, as suggested in \cite{Poland:2008ev}.  In the event that some Hyperbolic sfermions obtain vacuum expectation values, they are rendered unstable and decay back to the visible sector through the Higgs portal.

\section{Conclusions}
\label{sec:concl}
We have presented the first example of a neutral naturalness model with gauge neutral scalar top partners.  The key realisation was to convert the accidental $\U(4)$ symmetry of the Twin Higgs model into an accidental non-compact flat direction, which can be interpreted as an accidental $\U(2,2)$ symmetry of the scalar potential.  Critically, the one-loop quadratic corrections from visible sector tops and gauge neutral Hyperbolic stops preserve the flat direction and hence cancel at low energies.

After explaining the IR features of the mechanism, we then introduced a particular UV-completion set in a SUSY extra dimensional scenario with an exchange symmetry that enforced the equality of the top Yukawa couplings.  Scherk-Schwarz SUSY breaking boundary conditions for the matter fields were chosen to yield the required fermionic matter zero modes in the MSSM sector and scalar matter zero modes in the Hyperbolic sector.  A predominantly $\U(2,2)$-symmetric tree-level $D$-flat scalar potential is generated by a new $\U(1)_X$ gauge symmetry with an FI term.

It is useful to compare the Hyperbolic Higgs to other concrete realizations of neutral naturalness, namely the Twin Higgs and Folded SUSY. All three models feature $\eccoZ$ symmetries, albeit acting in different ways. In the case of the Hyperbolic Higgs, the symmetry acts only on the marginal couplings of the theory consistent with the $\U(2,2)$ accidental symmetry of the scalar potential, whereas the $\eccoZ$ of the Twin Higgs acts on all fields and the $\eccoZ$ of Folded SUSY acts strictly on the matter fields and $\SU(3)_c$ vector multiplets. The resulting accidental symmetry structure respected by the radiative corrections to the Higgs multiplets is $\U(2,2)$ in the case of the Hyperbolic Higgs, in contrast to the $\U(4)$ of the Twin Higgs and SUSY in the case of Folded SUSY. At the level of the IR effective theory, the top partners of the Hyperbolic Higgs are scalar and couple to the SM-like Higgs via effectively marginal interactions as in Folded SUSY (and in contrast to the Twin Higgs), but are entirely SM-neutral as in the Twin Higgs.

With regard to the gauge sector, there are significant differences between Hyperbolic, Twin, and Folded structures.  In the Hyperbolic Higgs one can choose SUSY-preserving boundary conditions in the SM and Hyperbolic electroweak gauge sectors to ameliorate cutoff sensitivity through gauge loops, as done here, or one could alternatively choose twisted boundary conditions in the Hyperbolic gauge sector such that the Hyperbolic gauge groups are broken but quantum corrections preserve the $\U(2,2)$ structure at the quadratic level.  However, this latter option suffers from the existence of additional unwanted pseudo-Goldstone bosons since the gauge bosons no longer eat the Goldstone bosons living in $H_\mathcal{H}$.

The most significant structural differences arise in the Higgs quartic interactions.  In the Twin Higgs model the quartic interactions and kinetic terms respect the $\U(4)$ symmetry, thus radiative corrections involving this quartic pose no risk to the mechanism.  However, in the Hyperbolic Higgs setup this is not the case, since $\U(2,2)$ is not a true symmetry of the Higgs sector and the quartic interactions lead to dangerous $\U(2,2)$-breaking sensitivity to the cutoff.  This is borne out in the UV-completion presented here, where the Higgs sector quartic interactions arise due to a new SUSY gauge force $\U(1)_X$.  SUSY must be broken to preserve the quartic coupling and remain consistent with precision electroweak bounds, but once SUSY is broken there is nothing to prevent the quartic couplings from generating $\U(2,2)$-breaking sensitivity to the SUSY breaking scale.  Indeed, this is what occurs in this UV-completion and presents the fundamental limitation to the efficacy of the model with respect to fine-tuning.  Thus, although the quadratic cutoff sensitivity due to the top sector is ameliorated effectively, a little hierarchy problem associated with the Higgs quartic interactions arises in its place.

As for phenomenology associated with the Hyperbolic Higgs mechanism, the SM-like Higgs is an admixture of SM and Hyperbolic Higgs doublets, leading to tree-level Higgs coupling deviations and a heavy radial mode of potential phenomenological relevance much as in the Twin Higgs. The fine-tuning of the IR effective theory is largely analogous to that of the Twin Higgs, though in the specific model considered here the tuning is instead dominated by constraints on the UV completion that may be avoided for other realizations. All three models share phenomenological signatures of dark QCD, though the Hyperbolic Higgs is most akin to Folded SUSY in that no $\eccoZ$-breaking is required for glueballs to dominate the low-energy spectrum.   However, for the Hyperbolic Higgs the production of Hyperbolic glueballs arises solely through the Higgs portal to Hyperbolic glue, whereas in Folded SUSY glueball production can additionally arise through Drell-Yan sparticle production. Overall,  within the class of neutral naturalness theories, the Hyperbolic Higgs appears as the example most resilient to collider scrutiny. This is because SUSY breaking effects in the Hyperbolic sector remove most of the light states below the electroweak scale. The only possible exceptions are Hyperbolic photons and gluons, which could also be massive if Hyperbolic color and electromagnetism are spontaneously broken by vevs of Hyperbolic scalar fields.

The way we formulated the Hyperbolic mechanism is independent of SUSY and one could imagine UV completions with composite Hyperbolic sectors or other variations. But even if SUSY does not originally participate in the definition of the Hyperbolic mechanism, it naturally embeds some of its characteristic features. There are three SUSY elements that are ideally suited for Hyperbolic implementations. First, scalar interactions with coupling constants related to the top Yukawa, as in \cref{eq:lag} and needed to preserve the $\U(2,2)$ symmetric structure of the potential after one-loop top corrections, appear naturally in a SUSY context. Second, SUSY gauge $D$-terms under which both the SM and the Hyperbolic Higgs are charged, can automatically give a quartic scalar interaction that is $\U(2,2)$ symmetric and of the form required by \cref{eq:scal}, as well as $\U(2,2)$ symmetric mass terms through the FI term. Third, the Hyperbolic mechanism can take care of the naturalness problem created by SM one-loop corrections from top Yukawa interactions and Higgs self-couplings, but needs assistance in the multi-TeV region to deal with quadratic divergences from the gauge sector. The SUSY completion of the Hyperbolic mechanism propitiously provides the missing element by regulating these divergences. 

In spite of this deep connection between Hyperbolic and SUSY, as far as collider searches are concerned, the Hyperbolic Higgs is completely different from regular SUSY models, since all the usual sparticles are pushed beyond the LHC discovery limits, at a mild fine-tuning price. Hopes to identify Hyperbolic features at colliders mostly rely on precision studies of Higgs properties, on exotic Higgs decays, or on the direct production of the Hyperbolic Higgs through its mixing with the ordinary Higgs -- the SM-like Higgs boson would provide the portal into the Hyperbolic sector.

\vspace{10pt}
\noindent \textbf{Note added:}  As this paper was being completed, we became aware of the complimentary work that will appear simultaneously~\cite{Cheng:2018gvu}.

\newpage

\acknowledgments

We wish to thank Kiel Howe for useful discussions and Antonio Delgado for pointing out an issue with our discussion of the Higgsinos.  TC is supported by the U.S. Department of Energy under grant number DE-SC0018191.  NC is supported by the U.S. Department of Energy under grant number DE-SC0014129.  MM, NC, and TC thank the CKC-CERN workshop ``What's going on at the weak scale?'' and MM and TC thank the Les Houches Workshop ``Physics at TeV Colliders'' 2017 for providing stimulating environments in which part of this work was developed.

\appendix
\section*{Appendices}
\section{Infrared Parameterization}
\label{app:ir}
Although we have focused on a specific SUSY UV completion of the Hyperbolic Higgs, it is useful to understand the parametrics of the Hyperbolic Higgs in terms of an IR effective description that can be matched on to diverse UV completions. The low-energy potential for the SM and Hyperbolic Higgs doublets can be put into the general form
\begin{align}
V = \lambda\Big(|\HH|^2 - |H|^2 - f^2\Big)^2 + \kappa\, \Big(|H|^4 + |\HH|^4\Big) + \rho\, |H|^4 + \sigma \,f^2\, |H|^2\,.
\end{align}
Here $\lambda$ respects the approximate $\U(2,2)$ symmetry of the theory, while $\kappa$ respects the $\eccoZ$ symmetry of the marginal couplings. The parameters $\rho$ and $\sigma$ constitute hard and soft breaking of the $\eccoZ$ symmetry. 

From here it is straightforward to understand the vacuum parametrics in the limit of $\lambda \gg \kappa, \rho$ and $f \gg v$. In this limit, the $\U(2,2)$ quartic $\lambda$ rigidly fixes $|\HH|^2 = f^2 + |H|^2$ on the Hyperbolic goldstone manifold, giving rise to the leading potential for the light doublet,
\begin{equation}
V \rightarrow \big(2\, \kappa + \rho\big) |H|^4 + \big(\sigma + 2\, \kappa\big) f^2\, |H|^2 \equiv - 2\, \lambda_\text{SM}\, v^2\, |H|^2 + \lambda_\text{SM}\, |H|^4\,.
\end{equation}
The only change relative to the compact $\U(4)$ case (as in \cite{Craig:2015pha}) is the flip of the sign of $\kappa$ in the mass-squared term. We can now express $v$ in terms of the underlying parameters via
\begin{equation}
\frac{2\, v^2}{f^2} = \frac{- \sigma - 2 \,\kappa}{2\, \kappa + \rho}\,,
\end{equation}
while the mass of the SM-like Higgs is
\begin{equation}
m_h^2 = 4\, \lambda_\text{SM}\, v^2 = 4 \big(2\, \kappa + \rho \big) v^2 = -2 \big(\sigma + 2\, \kappa \big) f^2\,.
\label{eq:IRmh}
\end{equation}
This provides clear conditions for both electroweak symmetry breaking and parametric separation $v \ll f$ in terms of the effective parameters of the Hyperbolic Higgs potential.

Much as in the Twin Higgs, in the goldstone approximation, we can compute the overall tuning of the electroweak scale as the product of two tunings: the tuning of the scale $f$, which is set by physics at the cutoff $\Lambda$; and the tuning of $v/f$, which is set by the symmetry-breaking terms in the potential,
\begin{equation}
\Delta_{v^2} = \Delta_{f^2} \times \Delta_{v^2/f^2}\,\,.
\end{equation}
Both tunings can be straightforwardly computed in terms of the IR parameterization with cutoff $\Lambda$, or in terms of the UV completion that fixes the IR parameters.

In order to map onto our UV parameters, we can rewrite \Eq{eq:IRmh} as
\begin{equation}
m_h^2 =   - \frac{\sigma + 2 \,\kappa}{\lambda} m^2 \,.
\end{equation}
Taking into account both $\U(2,2)$-breaking quartic couplings we find
\begin{equation}
m_h^2 = \frac{c_t \,\lambda_t^2- g_Z^2}{\lambda} m^2 +\dots \,\,,
\end{equation}
where the relevant loop factor is
\be
c_t = \frac{N_c\, \lambda_t^2}{48\, \pi^2 } \Big[ 8-6\, \log \alpha_{t_\mathcal{H}} +21\, \zeta(3) \Big] \,.
\ee
Thus, a naturally light SM-like Higgs requires $\lambda$ to be as large as possible.  Furthermore, recalling that $m$ can be determined from \Eq{eq:soft} and noting that the colored top partner fields and the stop squarks have mass $m_{\tilde{t}} = 1/(2\, R)$, we see that unless the FI-term is fine-tuned the quadratic sensitivity to the cutoff is given by
\begin{equation}
m_h^2 = \frac{c_t\, \lambda_t^2- g_Z^2}{\lambda}  \frac{21 \,\zeta (3)\, \lambda_t^2 N_c}{8\, \pi^4} m^2_{\tilde{t}} +\dots \,\,.
\end{equation}
As a result, we see that the SM Higgs boson mass is naturally a loop factor below the mass scale of colored stop squarks, with an \emph{additional} suppression factor proportional to $\sim g^2_{\text{SM}}/\lambda$, where $g_{\text{SM}}$ involves the electroweak gauge coupling or the top Yukawa.  This result is consistent with the general arguments applied to the Twin Higgs in \cite{Contino:2017moj}.

%\section{Santa Clausen's Expansion}
\section{Analytic Minimisation Conditions and the Higgs Mass}
\label{app:santaclausen}
In this section we provide useful exact expressions for the CW potential in \cref{eq:VCWsummed} and its derivatives, give analytic expansions and discuss the conditions under which these formulae provide good approximations.  

To begin, we define the Clausen functions, which are related to the $n^\text{th}$ polylogarithms $\text{Li}_n(z)$ as
\begin{align}
\text{Cl}_n (x)= 
\left\{
\begin{array}{cc}
\frac{i}{2}\left[\text{Li}_n\big(e^{-i\, x}\big)- \text{Li}_n \big(e^{i\, x}\big)\right] & \qquad n \text{ even;} \\[10pt]
\frac{1}{2}\left[\text{Li}_n\big(e^{-i\, x}\big)+ \text{Li}_n \big(e^{i\, x}\big)\right] & \qquad n \text{ odd.} 
\end{array}
\right. \,
\end{align}
They can be represented as series
\begin{align}
\text{Cl}_{2n}(x)=\sum_{k=1}^\infty \frac{\sin k\,x}{k^{2n}}\,;\qquad\qquad \text{Cl}_{2 n+1}(x)=\sum_{k=1}^\infty \frac{\cos k\,x}{k^{2 n+1}}\,,
\end{align}
and their derivatives are given by
\begin{align}
\frac{d\text{Cl}_{n+1} (x)}{dx}=(-)^{n+1}\text{Cl}_{n} (x)\,.
\end{align}

We then define the following functions\footnote{Equivalently, the functions $F_n$ can be written as $F_n(x) =2[\text{Cl}_n ( 2\, x  )-2^{-n} \, \text{Cl}_n ( 4\, x  )]$.}
\be
F_n (x)  =   \text{Cl}_n ( 2\, x  ) -\text{Cl}_n ( 2\, x +\pi ) \, .
\ee
Useful Taylor expansions are\footnote{These expansions are obtained by using the derivative rule
$$ \frac{d^p F_n(x)}{dx^p} = (-)^{p(2n-p+1)/2}\, 2^p\, F_{n-p}(x) ~~~{\rm for}~p\le n-1\, ,$$
while higher derivatives ($p>n-1$) can be computed by using $F_1(x) =-\log |\tan x|$. The values of $F_n$ at the expansion point are given by
$$F_n(0)=0 ~~{\rm for}~n~{\rm even},~~~F_n(0)=2(1-2^{-n})\zeta(n) ~~{\rm for}~n\ge 3~{\rm odd},~~~
F_1(x) \underset{x\to 0}{\longrightarrow} -\log |x| \, .$$
}
\begin{align}
F_3 (x ) & =  \frac{7}{4}\, \zeta(3)+x^2 \left(\log  x^2 -3  \right)+ \ord{(x^4)}\,, \\[5pt]
F_4 (x ) & =   \frac{7}{2}\, \zeta(3)\, x+ \frac{2}{9} \,x^3 \left( 3 \log  x^2  -11 \right) + \ord{(x^5)}\,, \\[5pt]
F_5 (x ) & =   \frac{31}{16}\, \zeta(5)-\frac{7}{2}\,\zeta(3)\, x^2+ \frac{x^4}{3} \left( \frac{25}{6} - \log  x^2 \right) + \ord{(x^6)}\,.
\end{align}

Taking $q_B=1/2$ and $q_F=0$, the CW potential in \Eq{eq:VCWsummed} becomes
\be
V_\text{CW} (H) =\frac{3\, N_c}{16\, \pi^6\, R^4}\,  F_5 \big( \pi\, R\, m_t(H) \big)    \,.
\ee
Its derivatives evaluated on the vacuum are
\begin{align}
\frac{\partial V_\text{CW} (H)}{\partial H} \bigg|_{H= v} &=-\frac{3\, N_c\, \lambda_t\, \cos^2 \alpha_t}{8\, \pi^5\, R^3}\, F_4 (\alpha_t)    \,,
\label{eq:expandedCW1} \\[10pt]
\frac{\partial^2 V_\text{CW} (H)}{\partial H^2} \bigg|_{H= v} &= \frac{3\, N_c\, \lambda_t^2\, \cos^4 \alpha_t}{4\, \pi^4\, R^2}  \Big[    \tan\alpha_t\, F_4 (\alpha_t) - F_3 (\alpha_t)\Big]    \,,
\label{eq:expandedCW2}
\end{align}
where we recall that $\alpha_t \equiv \pi\, R\, m_t (v)$.  The contributions to the CW potential from the Hyperbolic sector are found by making the replacement 
\be
V_\text{CW} (H_{\mathcal{H}}) = - \left. V_\text{CW} (H\to H_{\mathcal{H}})\right|_{\alpha_t \to \alpha_{t_\mathcal{H}}}\,.
\ee

Note that, for $\lambda_t\, v_{\mathcal{H}} \sim 1/(2\, \pi\, R)$ (a plausible size for the Hyperbolic Higgs vacuum expectation value) the second order correction in $F_3 (\alpha_{t_{\mathcal{H}}} )$ leads to a $50\%$ reduction relative to the leading order value, while the terms such as $\cos^4 \alpha_{t_{\mathcal{H}}}$ can lead to additional reductions of over $\sim 40\%$.  As a result, while it is always adequate to expand in $\alpha_t$, the perturbative expansion to lowest orders in $\alpha_{t_{\mathcal{H}}}$ is not always valid over all of the parameter space that we consider, so care must be taken when using the expanded expressions.

From the minimisation conditions of the scalar potential in \Eq{fullscalar} with respect to the Higgs and Hyperbolic Higgs, we can derive the required values of the soft mass and FI-term,
\bea
&\widetilde{m}^2 + \widetilde{m}_X^2   = \frac{1}{2\, \pi^2\, R^2} \left[ -\frac{g_Z^2}{\lambda_t^2}\Big(T_\mathcal{H}^2+T^2\Big)
+\frac{3\, N_c\, \lambda_t^2}{16\, \pi^2}\left( \frac{F_4(\alpha_t)}{T(1+T^2)} - \frac{F_4(\alpha_{t_\mathcal{H}})}{T_\mathcal{H}(1+T_\mathcal{H}^2)}\right) \right] \, ,\label{eqmin1} \\[15pt]
&f_{X}^2 = \frac{1}{2\, g_X^2\, \xi\, \pi^2\, R^2}\left[ \frac{2\, g_X^2\, \xi+g_Z^2}{\lambda_t^2}\Big(T_\mathcal{H}^2-T^2\Big)
+\frac{3\, N_c\, \lambda_t^2}{16\, \pi^2}\left( \frac{F_4(\alpha_t)}{T(1+T^2)} + \frac{F_4(\alpha_{t_\mathcal{H}})}{T_\mathcal{H}(1+T_\mathcal{H}^2)}\right) \right] \, ,
\label{eqmin2}
\eea 
where $T\equiv \tan \alpha_t=\pi\, R\,\lambda_t\, v$ and $T_\mathcal{H}\equiv \tan \alpha_{t_\mathcal{H}}=\pi\, R\,\lambda_t \,v_\mathcal{H}$.

The entries of the $2\times 2$ mass matrix, which mixes the Higgs and Hyperbolic Higgs scalars, are given by
\bea
 \mathcal{M}^2_{hh}  &=& 
2\big(g_X^2\, \xi + g_Z^2\big) v^2 - \frac{N_c\, \lambda_t^4\, v^2\, G(T)}{4\,\pi^2} \, ,
\nonumber \\[10pt]
  \mathcal{M}^2_{h_\mathcal{H}h_\mathcal{H}}  &=& 
2\big(g_X^2\, \xi + g_Z^2\big) v_\mathcal{H}^2 + \frac{N_c\, \lambda_t^4\,v_\mathcal{H}^2\, G(T_\mathcal{H})}{4\,\pi^2} \, ,
 \\[10pt]
 \mathcal{M}^2_{hh_\mathcal{H}}  &=&   -2\, g_X^2\, \xi\, v\, v_\mathcal{H}\nonumber\,,
\eea
where
\bea
G(x) & = & \frac{3}{2\,x^2(1+x^2)^2}\left[ F_3(\arctan x) - \frac{1+3\,x^2}{2x}\,F_4(\arctan x) \right] \\[10pt]
& = &  \log x^2 -c +x^2 \left( \frac{23}{10}\, c +\frac{16}{3}  - 4 \log x^2 \right) +{\mathcal O} (x^4) ~, \nonumber
\label{eq:GandExp}
\eea
and
\beq
c= \frac83 + 7\, \zeta(3) \simeq 11.08 \, .
\eeq
Note that in the Hyperbolic limit, defined in \Eq{eq:hyplim}, the determinant of $\mathcal{M}^2$ vanishes, as expected.  

In the limit $v\ll v_\mathcal{H}$, the lighter (SM-like) Higgs boson has mass
\beq
m_h^2 = (2-\Omega) M_Z^2 + \frac{N_c\, \lambda_t^4 }{2\, \pi^2}\, v^2 \log \frac{v_\mathcal{H}}{v} 
+ {\mathcal O}\left( \frac{v^4}{v_\mathcal{H}^2}\right) \,,
\label{massahig}
\eeq
where
\be
\Omega = \frac{ 
\frac{g_Z^2}{g_X^2\,\xi}  \big[ 1 +\kappa_t \, G(T_\mathcal{H}) \big] \big[ 1 +\kappa_t  (\log T_{\mathcal{H}}^2-c) \big]    -\kappa_t \big[ G(T_\mathcal{H})-\log T_{\mathcal{H}}^2 +c \big]}{1+ \frac{g_Z^2}{g_X^2\,\xi}\big[ 1+ \kappa_t \, G(T_\mathcal{H}) \big]} 
\label{eq:Omega}
\ee
and
\beq
\kappa_t = \frac{N_c\, \lambda_t^4}{8\,\pi^2\, g_Z^2}  \, .
\eeq
In \cref{massahig} we have explicitly separated the contribution from the UV matching onto the SM effective theory (parametrized by the coefficient $\Omega$) and the model-independent IR contribution due to the running within the SM effective theory from the new-physics scale to the weak scale (captured by the term proportional to $(\lambda_t^4/\pi^2)\log v$). 

Note that the coefficient $\Omega$ goes to zero as we approach the Hyperbolic regime ($g_Z^2\ll g_X^2$), up to small $\kappa_t$ corrections (which vanish for small $R\,v_\mathcal{H}$). On the other hand, in the opposite limit of negligible $g_X$ and $\kappa_t$, we find $\Omega =1$, recovering the familiar UV matching condition of minimal low-energy SUSY in the large $\tan\beta$ limit. 

The coefficient $\Omega$ also describes the dependence of the Higgs mass on the compactification radius $R$. For $g_Z^2/ g_X^2\ll 1$
and $Rv_\mathcal{H} \ll 1$, we find
\beq
\Omega = \frac{N_c\, \lambda_t^6\, R^2 \,v_\mathcal{H}^2}{g_Z^2} \left[ \log (\pi\, \lambda_t\, R\, v_\mathcal{H})
-\frac{23}{80}\, c-\frac23 \right]  +{\mathcal O} \left( \frac{g_Z^2}{g_X^2}, R^4\,v_\mathcal{H}^4 \right)\, .
\eeq
We recall that the above approximation is not always valid in the entire parameter space of the theory, in which case the exact expression of $\Omega$ in \Eq{eq:Omega} must be used.

The mixing angle $\theta$ which defines the SM-like eigenstate $h_{\text{SM}}$ according to \Eq{eq:mixing} is given by
\beq
\sin \theta = \omega \, \frac{v}{v_\mathcal{H}} +
{\mathcal O}\left( \frac{v^3}{v_\mathcal{H}^3}\right) \, .
\eeq
\beq
\omega =\frac{1}{1+ \frac{g_Z^2}{g_X^2\,\xi}\Big[ 1+ \kappa_t\,  G\big(T_\mathcal{H}\big) \Big]}
\label{omeghino}
\eeq
The heavy (mostly Hyperbolic) Higgs boson has mass
\beq
m_H^2 = \frac{2\, g_X^2\, \xi \, v_\mathcal{H}^2}{\omega} \left[ 1+ {\mathcal O}\left( \frac{v^2}{v_\mathcal{H}^2}\right) \right]\, .
\eeq

\bibliographystyle{utphys}
\bibliography{HyperbolicHiggs}

\end{document}